\documentclass[english]{revtex4}
\usepackage{graphicx}
\usepackage{amsmath}
\usepackage{bm}
\usepackage[english]{babel}
\begin{document}

\title{Time dependence of X-ray diffraction intensity of a crystal induced by an intense femtosecond X-ray pulse}

\author{A. Leonov$^1$, D. Ksenzov$^2$, A. Benediktovitch$^1$, I. Feranchuk$^1$ and U. Pietsch$^2$}

\address{$^1$Department of Theoretical Physics, Belarusian State University, 220030 Nezavisimosti ave. 4, Minsk, Belarus \\
$^2$ Festk\"{o}rperphysik, Universit\"{a}t Siegen, 57072 Walter-Flex-Str. 3, Siegen, Germany}

\begin{abstract}
\noindent The time evolution of the electron density   and the resulting time dependence of X-ray diffraction peak intensity in a crystal irradiated by highly intense femtosecond pulses of an XFEL is investigated theoretically on the basis of rate equations for bound electrons and the Boltzmann equation for the kinetics of the  unbound electron gas   that plays an essential role in the time evolution of the electron density of a crystal. The photoionization, Auger process, electron-impact ionization, electron--electron scattering, and three-body recombination have been implemented in the system of rate equations.  An algorithm for the numerical solution of the rate equations was simplified  by incorporating analytical expressions for the cross sections of all the electron configurations in ions   within the framework of the effective charge  model. Using this approach we evaluate the time dependence of the inner shell population and electronic kinetic energy during the time of XFEL pulse propagation through the crystal for photon energies between 3 and 12 keV and a pulse width of 40 fs considering a flux of $10^{12}$\ ph/pulse (focusing on a spot size of $ \sim 1 \mu$m$^2$, this flux corresponds to a fluence ranging between 0.6 and 1.6 \ mJ/$\mu$m$^2$). The time evolution of the atomic scattering factor and its fluctuation is numerically analyzed for the case of a Silicon crystal taking into account the decrease of the bound electron density during the pulse propagation.   The time integrated intensity drops dramatically if the fluence of the XFEL pulse exceeds 1.6 mJ/$\mu$m$^2$.
\end{abstract}

\maketitle

PACS number(s): 32.80.Fb, 32.90.+a, 87.59..e, 87.15.ht

\section{Introduction}  The first hard X-ray Free Electron Lasers (XFEL) \cite{XFEL1, XFEL2, Chapman_2009Nat} are already in operation at SLAC (USA) and Spring-8 (Japan); other XFEL facilities are under construction, including the European XFEL at DESY \cite{EuroXFEL_tech}. These facilities will provide ultra-bright femtosecond X-ray radiation with unique possibilities to study the structure of matter with angstrom resolution on a time scale of femtoseconds. Most of the current experiments using FEL radiation focus on single shot exposure of molecules and clusters, assuming that structure data can be taken before sample destruction takes place \cite{image} on a time scale much larger than the FEL pulse length. Having this sample destruction in mind, FEL experiments on crystals are rare at present \cite{Shastri, PSI2009}. Specific experimental conditions for FEL experiments have to be defined in order to solve specific questions of solid state physics.

At present, crystal diffraction is used for monochromators or other optical elements. During the first experiments with XFEL sources it was discovered that the crystal response known from conventional experiments at synchrotron sources is maintained as long as the fluence, i.e., the deposited photon energy per sample area, is below a certain threshold \cite{Hau-Riege, Hau-Riege-LCLS}. Therefore for current experiments the crystal is illuminated by a wide beam and the focusing takes place after monochromatization. However, other experimental scenarios might be realized in future experiments. One is the photon--photon pump-probe experiment where the sample is excited by one FEL pulse followed by a second one after a time span much shorter than the repetition time of the FEL source. A respective time delay setup has been proposed recently, equipped with four crystal reflections \cite{Roseker}. For this experiment it is important to know how both the pulse shape and the intensity of the delayed pulse differ from those of the first pulse if a highly intense FEL fs-pulse propagates throughout the crystal \cite{Bushuev, Shvydko}.
The answer to these questions may also influence the design of other optical elements to be positioned within the path of the XFEL beam. Up to now the interaction of FEL pulses with a crystal has been described by many authors in terms of X-ray dynamical theory considering the time delay of the X-ray beam while propagating through the crystal \cite{Shvydko, Malgrange} but using time independent atomic scattering factors (ASF). However, it was shown in \cite{Hau-Riege-book} that such an approach remains valid only in case of relatively small fluences. In our paper we will show that the major variation in diffraction intensity originates from the alteration of the ASF due to electronic processes. As long as crystal diffraction can be described in terms of kinematic theory, the relevant quantities for the description of the integrated intensity are the ASF and its variation as a function of time and fluence.

In the fs time range, the atomic positions in a crystal are fixed and the main source of variation is the electronic excitation and Auger recombination of bound electrons induced by the X-ray beam. Because the time scale of these processes is in the same time range as the FEL pulse length, the population of electronic states of an atom and subsequently the atomic form factor become time dependent. Under these conditions, conventional theories of X-ray diffraction that are based on the stationary X-ray susceptibility of the crystal \cite{Authier} are no longer valid   because of the fast evolution of the electron density in the crystal. Since the duration $\tau_D$ of the formation of a diffraction peak, defined by the extinction length $L_{ext}$ ($\tau_D \approx L_{ext}/c \sim 10 \ fs$, where $c$ is the speed of light) is comparable with the duration of the XFEL pulse it is necessary to take into account the dynamics of electronic redistribution within the atomic shells. These processes finally result in the time dependence of the ASF and the integrated Bragg peak intensity.

The evolution of  electron density of an object irradiated by an XFEL pulse can be described by the solution of rate equations for the atomic state populations (e.g. \cite{Santra-Son},  \cite{SantraPhoto} and references therein) or by the simulation of microscopic processes in terms of the Monte Carlo method \cite{Hau-Riege-book}. An alternative approach is focussing on description of evolution of the electron plasma that is created in the process of ionizing the atoms (e.g. \cite{ZiajaNew}, \cite{Hau-Riege2013} and references therein). Moreover, it was also shown \cite{Rost2012}--\cite{Schorb} that the ionization dynamics of individual atoms changes substantially considering the influence  of the electron plasma on the time dependent evolution of the population probabilities. As a result, the population of the atomic configurations depends on the relation between pulse duration and the size of the cluster on the one hand and the energy  distribution of plasma electrons on other hand \cite{Schorb}.  Evidently, the latter effect becomes essential in the case of crystals where the electronic band spectrum differs substantially from the energy spectrum of electrons in isolated atoms and molecules.

Specific feature of our approach is based on the numerical solution of a self-consistent system of master equations that includes both the rate equations for the population of bound electrons and the Boltzmann kinetic equation for the distribution function of unbound (plasma) electrons generated by the ionization of the atoms during the pulse propagation in the medium. Such an approach allows one i) to trace explicitly evolution of all possible atomic/ionic configurations as it is vital for further estimation of the X-ray diffraction intensities (this means that if one considers an ion with total charge +1 the diffraction signal is different for the cases of inner and outer vacancies) and ii) to take into account secondary ionization processes and the role of free electron plasma in the problem of evolution of atomic states of the system. The latter part considers the band spectrum of unbound electrons and additional relaxation, such as the ionization of the atoms by the electrons, electron--electron collisions, and three-body recombination. The numerical treatment of these additional processes makes the solution of the master equations very expensive. Therefore  analytical expressions for the cross sections of all the electron configurations in the ions have been derived on the basis of the effective charge model (ECM) for single-particle atomic wave functions \cite{Acta, FerTriguk}. They have been implemented in the numerical algorithm of the solution of the master equations. The developed software, ``crystal evolution induced by X-ray'' (CEIX),  is applicable to various atoms. Its possibilities are demonstrated for a Si crystal as the example.

The present paper deals with theoretical investigation of the electron density evolution of atoms arranged in a crystal and the estimation of the time dependence of the Bragg peak intensities during the propagation of an intense XFEL fs-pulse through the crystal. As shown in \cite{Santra-Ziaja2012}, the ASF decreases remarkably during the time of the pulse propagation through the sample. This means that the conventional linear theory of diffraction, assuming a constant crystal susceptibility  is no longer valid \cite{Shvydko}.

We concentrate on the calculation of the population dynamics of the atomic electronic states considering bound and unbound electronic states and the resulting time dependence of the ASF. The time dependence of the Bragg peak intensities is estimated from the square of the structure factors making up the time dependent ASFs. Bragg peak intensities are described in terms of kinematical theory of X-ray diffraction which is valid as long as the crystal thickness is smaller than the extinction length $L < L_{ext} \approx c/\omega|\chi_g|$ (where $\chi_g$ is the Fourier component of the X-ray polarizability of the crystal and $\omega$ is the frequency of X-ray radiation) so that the dynamical effects are negligibly small. For Silicon at 8 keV photon energy $L_{ext} = 18.5$ $\mu$m at (111) reflection in Laue geometry, for instance \cite{Stepanov}.

Considering its fs time range, the FEL pulse will probe a snapshot of the atomic arrangement in the crystal affected by random displacements of the atoms due to thermal displacements. We suppose that the respective damping of the diffraction intensity can be effectively described in terms of the static Debye--Waller approach, causing a certain reduction in the Bragg peak intensity. Whereas this part is not considered in our approach for now, we effectively describe the evolution of the Bragg peak intensity by considering five different processes of electron redistribution in the atoms and their contributions to the ASF. The degree of electron redistribution depends on the pulse length and the pulse intensity, and becomes essential if the time necessary for complete ionization of the atoms is on the order of the time necessary to form the diffraction peak. We show results of numerical investigations at photon energies of 3 keV and 8  keV, i.e., close and apart from the Si K-edge,  using a pulse length of 40 fs and a flux of $10^{12}$\ ph/pulse (the fluence being $0.6$ and 1.6 mJ/$\mu$m$^2$ correspondingly).

The present paper is organized as follows. Secs. \ref{sec:2} and \ref{sec:3} motivate the approximations and introduce the processes considered for the description of the evolution of the electron density during the propagation of an XFEL pulse through a crystal. The complete system of master equations that describes the ionization dynamics in the crystal and the algorithm of the numerical solution are described in Sec. \ref{sec:4}. The numerical results for the evolution of the electron density are discussed in Sec. \ref{sec:5} followed by a description of the time dependence of the diffraction intensities from a Si  crystal described in Secs. \ref{sec:6} and \ref{sec:7}.

\section{Qualitative analysis}
\label{sec:2}
In general, the problem of the propagation of an X-ray pulse through matter is based on the solution of the system of Maxwell equations for the X-ray wave field coupled to the Schr\"odinger equation for the quantum states of the electron subsystem of the crystal. In contrast to the widespread approximation of linear X-ray optics that treats the electrons as classical oscillators \cite{Authier}, a quantum theory approach for the electron density response is required in order to take into account the variations of the atomic state populations during the interaction between the X-ray field and the crystal.

First of all, let us  estimate the effect of an intense X-ray laser field on a single atom using the parameters of the XFEL pulse introduced in the EuroXFEL technical design report \cite{EuroXFEL_tech}.

The electric field strength in the photon pulse can be evaluated as \cite{LandauV4}

\begin{eqnarray}
\label{1_1}
\mathcal{E}\approx \sqrt{\frac{4 \hbar \omega N_{ph}}{\epsilon_0 \pi d^2 c T}} \sim 4 \times 10^{9} \frac{V}{m} <   \mathcal{E}_a   \approx 5 \times 10^{11}\frac{V}{m},
\end{eqnarray}

\noindent where $\epsilon_0$ is the dielectric constant,   $\mathcal{E}_a = m^2 c^3 \alpha^3/ (\hbar e_0) $ is the characteristic strength of the atomic field with $\alpha$ being fine structure constant; $e_0$ is the electron charge.

The effect of an alternating laser field on the non-resonant atomic states is defined by the ponderomotive energy \cite{Popov}

\begin{eqnarray}
\label{2_1}
U_p \approx \frac{e_0^2 \mathcal{E}^2}{2 m \omega^2 }  \sim 2 \times    10^{-9} \ eV,
\end{eqnarray}

\noindent that is essentially smaller than the average atomic ionization potential $U_i$. The probability of non-resonant ionization of atoms by a laser field can be  calculated on the basis of \cite{Popov}. In the considered case the Keldysh parameter

\begin{eqnarray}
\label{2_2}
    \gamma = \sqrt{ U_i/2 U_p } \gg 1,
\end{eqnarray}

\noindent which means that the probability of under-barrier tunneling is extremely small.

Taking into account (\ref{1_1})--(\ref{2_2}), one can conclude that the atomic wave functions represent a good basis set to describe the atom--field interaction in terms of perturbation theory.

\begin{figure}[t]
\includegraphics[scale=0.5]{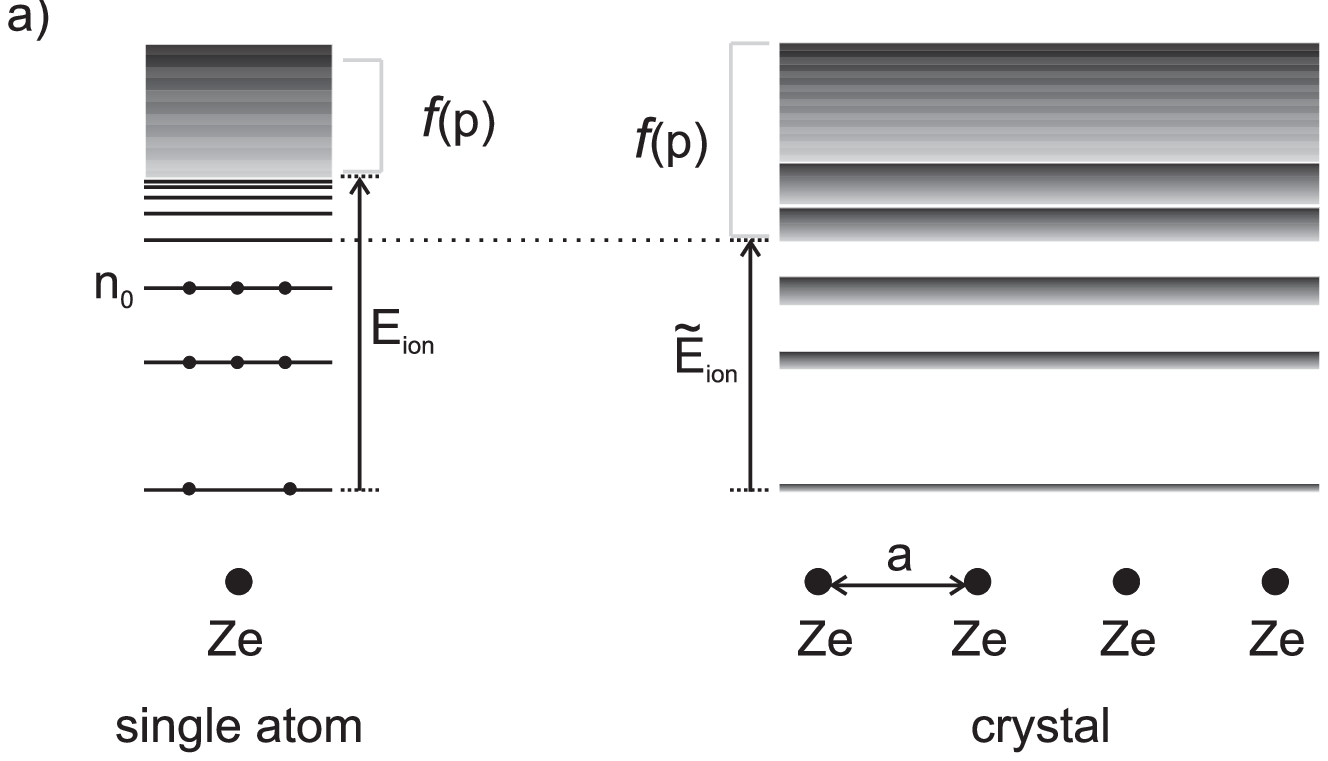}
\, \, \, \, \, \, \, \,
\includegraphics[scale=0.5]{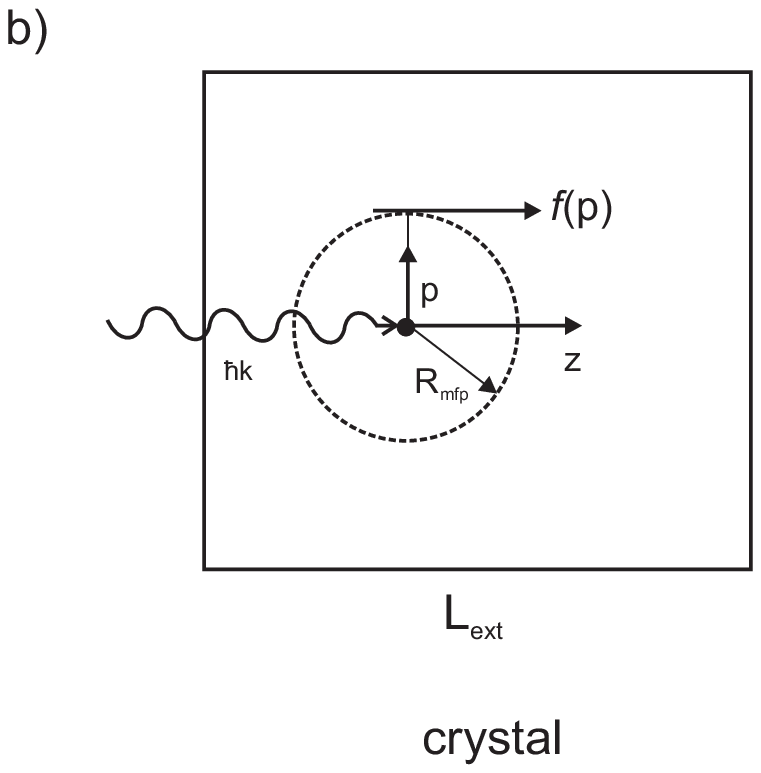}
\caption{(a) comparison of energy spectra of the electron states in isolated atoms and atoms in a crystal; (b) schematic estimation of the role of free electrons in the ionization dynamics. Here $f(p)$ is the distribution function of the free electrons, $R_{mfp}$ is the electron mean free path, and $L_{ext}$ is the extinction length.}
\label{Fig.1}
\end{figure}

Let us compare the typical structure of the energy spectrum of electron states in a crystal \cite{solid} with the energy spectrum of a single atom (Fig. \ref{Fig.1}a). The overlap of the electron shells of the atoms in the crystal leads to the formation of the energy bands $\tilde{E}_n(\vec p)$ ($n$ is the zone number, $\vec p$ is the quasi-momentum). The electron states with  $n \leq n_0$ ($n_0$ is the quantum number of the highest populated energy level for bound electrons) correspond to the ground state of the system, the widths of the allowed bands are defined by the exponentially small overlap integrals between neighboring atomic states \cite{solid}, so that the energy levels in every unit cell are approximately equal to the $\tilde{E}_n(\vec p) \approx E_n$, found at an isolated atom. At the same time, the excited states with $n > n_0$ correspond to the conduction band. For these states the overlap integral is large and the energy spectrum is described in the framework of the ``free electron approximation'' \cite{solid} by $\tilde{E}_n(\vec p) \approx p^2/2m$. This behavior is opposite to the case of an atom in a molecule or a small cluster, where the energy of the unoccupied states is still sharp.  Due to the formation of the band structure the effective ionization energy that defines the transition of the electrons from the discrete to the continuous spectrum becomes a little bit smaller in a crystal than in a molecular system.

Another important feature of the ionization dynamics in crystals is the role of the free electrons, which are described by the distribution function $f(\vec p)$ (Fig. \ref{Fig.1}b). The characteristic energy of the free electrons that appear due to the photoionization is defined by the photon energy $ p^2/2m \simeq  \hbar \omega \sim 10$\ keV.  The mean free path $R_{mfp}$ of the electrons of such an energy in media is defined by the energy loss due to secondary ionization processes, and according to the NIST database \cite{NIST} it can be estimated as $R_{mfp} \sim 10$ nm. At the same time, in a crystal with a thickness of the same order of magnitude as the extinction length $L \sim L_{ext}$ the percentage of ionized electrons that remain within the crystal can be approximately estimated as:

\begin{eqnarray}
\label{3a}
\lambda \approx \left( 1 - \frac{R_{mfp}}{L_{ext}} \right) \simeq 1,
\end{eqnarray}

\noindent that is almost a unity. This means that in a crystal, a considerable part of the free electrons contributes to the evolution of the electron density.

\begin{figure}[tb]
\includegraphics[scale=0.75]{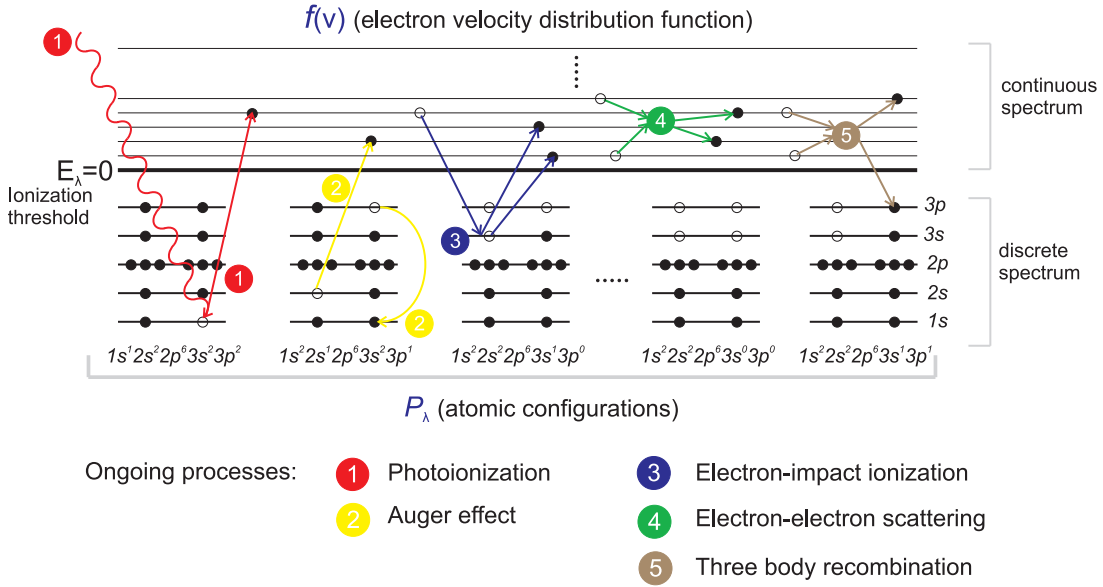}
\caption{ Elementary processes that define ionization dynamics in the crystal.}
\label{Fig.3}
\end{figure}

\section{Basic assumptions and justifications}
\label{sec:3}

The contribution of free electrons to the redistribution of the electron density is essential and needs to take into account additional elementary processes in order to define the ionization dynamics during the interaction of the XFEL pulse with the crystal (Fig. \ref{Fig.3}).  The interaction of the XFEL pulse with a single atom is described by photoionization and Auger processes \cite{Santra-Son} (processes 1 and 2, respectively). In a crystal, the large number of electrons excited into the conduction band leads to electron--electron collisions, electron impact ionization of other atoms, and the reverse process of a three-body recombination (processes 3 to 5, respectively). A sixth process is the possibility of induced photorecombination (not shown in Fig. \ref{Fig.3}). This process is reverse to photoionization, and takes place if the free electrons of the continuous spectrum become excited into unoccupied atomic states under the influence of the electromagnetic field pulse. This process is substantially resonant and involves free electrons with momenta $p_r \approx \sqrt{2m (\hbar \omega - E_n)}$. However, numerical results show (see Sec.\ref{sec:5} below) that due to the collisions with electrons and atoms, the photoelectrons quickly fill the entire range of the continuous states (Fig. \ref{picnum3}) and, hence, the contribution of the resonant photorecombination to the kinetic equation for the distribution function $f(p)$ can be neglected.

In order to find the intensity of a Bragg peak formed by the XFEL pulse, one has to calculate the crystal X-ray susceptibility taking into account the evolution of the electron density. Following textbooks as \cite{Landau8} one has to solve the Maxwell equations for the photon field (here the Coulomb gauge is used)

\begin{eqnarray}
\label{4 }
\nabla^2 \vec{A}(\vec{r},t)-\frac{1}{c^2}\frac{\partial^2 }{\partial t^2}\vec{A}(\vec{r},t)=-\frac{4\pi}{c}\frac{\partial \vec{j}(\vec{r},t)}{\partial t},\nonumber\\   \nabla \vec{A}=0, \varphi=0,
\end{eqnarray}

\noindent with $\vec{A}$ and $\varphi$ being the vector and scalar potentials correspondingly, coupled to the Schr\"odinger equation for the wave functions $\Psi_a (\vec r, t) = \Psi (\vec r - \vec R_a, t)$ of the electron subsystem of  the atom in the crystal unit cell localized near the  point $\vec R_a$:

\begin{eqnarray}
\label{5}
i\hbar \frac{\partial \Psi_a(\vec{r},t)}{\partial t}=\hat{H}\Psi_a(\vec{r},t),\nonumber\\
 \hat{H}=\frac{1}{2 m}\left( \hat{\vec{p}}-\frac{e_0}{c}\vec{A}(\vec{r},t) \right) ^2 + e_0 V(\vec{r}),
\end{eqnarray}

\noindent where $V(\vec{r})$ is the part of periodic potential of the crystal within the considered unit cell.

The induced current density in the matter can be calculated as the sum over all cells:

\begin{eqnarray}
\label{6}
\vec{j}(\vec{r},t)= \sum_a \left\{ \frac{i e_0 \hbar}{2 m}(\nabla \Psi_a^* \left( \vec{r},t)\Psi_a(\vec{r},t)- \right.\right. \nonumber\\
\left. \Psi_a^*(\vec{r},t)\nabla\Psi_a(\vec{r},t)\right)-  \left. \frac{e^2}{m c}
\vec{A}(\vec{r},t)\Psi_a^*(\vec{r},t)\Psi_a(\vec{r},t) \right\}.
\end{eqnarray}

According to the analysis mentioned above, the stationary single-electron wave functions $\psi_n(\vec{r})$ of the electrons in the crystal can be used as a basis set for the solution of Equation (\ref{5}). Let us consider the evolution of the electron state with the quantum number $l$ and expand the wave function as follows:

\begin{eqnarray}
\label{7}
\Psi_a (\vec{r},t) = C_l \psi_l(\vec{r}-\vec{R}_a) +   \sum_{n\neq l} C_n \psi_n(\vec{r}-\vec{R}_a); \nonumber\\
C_{l,n} \equiv C_{l,n}(\vec{R}_a,t)= a_{l,n}(\vec{R}_a,t)e^{- \frac{i}{\hbar} E_{l,n} t };
\nonumber\\  \vec{A}(\vec{r},t)=\vec{A}_s(\vec{r},t)e^{i(\vec{k} \vec{r}-\omega t)}+c.c.
\end{eqnarray}

The quantum number $n$   corresponds to the entire set of the single-electron quantum states including the wave functions of the  continuous spectrum.  The coefficients $a_{l,n}(\vec{R}_a,t)$ and the slope functions $\vec{A}_s(\vec{r},t)$ (temporal envelope of the pulse) \cite{Santra-Ziaja2012} are varying   due to the atom--field interaction rather slowly in comparison with the atomic frequencies.

In the numerical calculations below, the analytical single-electron approximation---ECM  \cite{Acta, FerTriguk}---is used both for the functions $\psi_{l,n}(\vec{r})$ and the energies $E_{l,n}$ of the atomic stationary states. This approximation is based on the use of hydrogen-like wave functions with an effective charge for each orbital so that it provides an accuracy comparable to the results obtained by the Hartree--Fock approximation \cite{LANL}.

The conventional approach of calculating the linear response of a system (susceptibility) \cite{Batterman} is based on the approximation $a_l = 1$ and $a_n(t)$ being calculated by means of the perturbative solution of Equation (\ref{5}). In the present case, a lot of atomic transitions are excited at the same time due to the very strong field. This results in a significant depopulation of the initial state, which must be taken into account when calculating the non-linear and time-dependent response. If one neglects the transitions between different excited states during the pulse propagation (we assume these states to be located in the continuous spectrum), a compact equation for the function $a_n(t)$ can be derived:

\begin{eqnarray}
\label{8aY}
a_n(\vec{R}_a,t) =-i\frac{e_0}{2m c}\int_{-\infty}^{t}dt'\vec{A}_s(\vec{R}_a,t')a_0(\vec{R}_a,t')\nonumber\\
\times \left\langle \Psi_n(\vec{\rho}) \left| \hat{\vec{p}}e^{-i\vec{k} \vec{\rho} }\right| \Psi_0(\vec{\rho})\right\rangle e^{i(\omega_{n0}-\omega)t'}; \ \vec \rho = \vec r - \vec R_a; \ n \neq l;\nonumber\\
\dot{a_l}(\vec{R}_a,t) =-\frac{e_0^2}{4 m^2 c^2}\vec{A}^*_s(\vec{R}_a,t)\nonumber\\
\times \int_{-\infty}^{t}dt'e^{i \omega (t-t')}Y(t-t')\vec{A}_s(\vec{R}_a,t'){a_l}(\vec{R}_a,t'),
\end{eqnarray}

\noindent with the response function $Y(t-t')$, which allows one to take into account the effects of memory and coherence in the atom-field interaction:

\begin{eqnarray}
\label{9Y}
Y(t-t')=\sum_n \left\langle \Psi_0(\vec{\rho}) \left| \hat{\vec{p}}e^{i\vec{k} \vec{\rho} }\right| \Psi_n(\vec{\rho})\right\rangle \nonumber\\
\times \left\langle \Psi_n(\vec{\rho'}) \left| \hat{\vec{p}}e^{-i\vec{k} \vec{\rho}' }\right| \Psi_0(\vec{\rho}')\right\rangle e^{-i\omega_{n0}(t-t')}.
\end{eqnarray}

The resonant and non-resonant parts should be treated separately when solving Equation (\ref{8aY}) for $a_l(t)$. It can be shown that in the non-resonant case ($\omega_{nl} \neq \omega$)  the kernel of the integral operator (\ref{9Y}) is almost local in time because of the condition $\omega T \gg 1$. Then the decrease of population of the atomic ground state reduces to the rate equation

\begin{eqnarray}
\label{10rateeq}
\dot{a}_l (\vec{R}_a,t) =-I(\vec{R}_a,t) \sigma^{(tot)}(\omega) a_l(\vec{R}_a,t),
\end{eqnarray}

\noindent where $I(\vec {R}_a,t)$ is the XFEL field intensity at the point $\vec R_a$ of the considered atom and $\sigma^{(tot)}(\omega)$ is the total cross-section of inelastic scattering of the radiation by the atom. This value can be found experimentally by measuring the intensity dependent absorption coefficients $\mu = n_{res} \sigma^{(tot)}(\omega)$ ($n_{res}$ is the resonant atom  density)

Another approximation is used in the resonant case when $\omega_{l n_r} \approx \omega$ for one of the   transitions. Then the coupled equations define the populations of the resonant levels

\begin{eqnarray}
\label{11rabi}
i\dot{a}_l(\vec{R}_a,t)=- \Delta \omega  {a_l}(\vec{R}_a,t)+ U(\vec{R}_a,t)a_{n_r}(\vec{R}_a,t), \nonumber\\
i\dot{a}_{n_r}(\vec{R}_a,t) = - i\frac{\Gamma}{2}a_{n_r}(\vec{R}_a,t)+ U(\vec{R}_a,t)a_l(\vec{R}_a,t),
\end{eqnarray}

\noindent where $\Delta \omega = \omega -\omega_{l n_r}$, $\Gamma$ is the width of the excited level and $U(\vec{R}_a,t)$ is the coupling function defined as follows:

$$
U(\vec{R}_a,t) = -\frac{e_0}{2m c}\vec{A}_s(\vec{R}_a,t)\langle \Psi_{n_r}(\vec{\rho}) | \hat{\vec{p}}e^{-i\vec{k} \vec{\rho} }| \Psi_l(\vec{\rho})\rangle.
$$

Substituting Eq. (\ref{7}) in Eq. (\ref{6}) and summing over the periodic coordinates $\vec R_a$ of the crystal, one can find that the induced current includes only  the Fourier component  corresponding to set of the reciprocal lattice vectors $\vec h$:

\begin{eqnarray}
\label{12curr}
\vec{j} (\vec{r},t)=\sum_{\vec h} \vec{j}^{\vec{h}}_s(\vec{r},t)e^{i(\vec{k}+\vec{h})\vec{r}-\omega t}; \nonumber\\ \vec{j}^{\vec{h}}_s (\vec r,t) =
\frac{e_0}{m \Omega} \{ a_0^*(\vec{r},t)a_l(\vec{r},t)
  \langle \Psi_l(\vec{\rho}) | \hat{\vec{p}}e^{i(\vec{k}+\vec{h}) \vec{\rho} }| \Psi_l(\vec{\rho})\rangle  \nonumber\\
   - \frac{e_0}{  c  }\vec{A}_s(\vec{r},t) \sum_{l \leq l_m,j} |a_l(\vec{r},t)|^2 F_{l,j}(\vec{h})e^{i \vec h \vec R_j} e^{-W(h)} \},
\end{eqnarray}

\noindent where $\Omega$ is the unit cell volume and $F_{l,j}(\vec{h})$ is the partial atomic scattering factor that corresponds to the transferred scattering vector $\vec q = \vec h$. It is calculated for the state $\psi_l(\vec{r})$ and the coordinates $\vec R_j$   correspond to various atoms   in the unit cell, $e^{-W(h)}$ is the Debye--Waller factor \cite{Batterman}. The sum is calculated over all atoms within the crystal unit cell and all bound electron states with quantum numbers $l \leq l_m$ that were occupied in the initial state of the system. So the total scattering factor of the crystal unit cell is defined as follows

\begin{eqnarray}
\label{12a}
F(\vec h, t) = \sum_{l \leq l_m,j} |a_l(\vec{r},t)|^2 F_{l,j}(\vec{h})e^{i \vec h \vec R_j}.
\end{eqnarray}

Far from resonance, i.e., far from the K- or L-absorption edges, the anomalous dispersion term in (\ref{12curr}) can be neglected \cite{Kissel}. This means that only the last term in the induced current density (\ref{12curr}) defines the diffraction intensity.

The main processes that determine the dynamics of the occupation probabilities and the time-dependence of the current density via Eq. (\ref{12curr}) are the  photoionization and the Auger effect. Here we assume that the ionized electrons are described by plane waves and do not contribute to the periodic susceptibility. However they can strongly affect the bound electron population. During the pulse propagation the inner shells become depleted due to both photon-induced processes and electron--atom impact ionization.

\section{Application of the rate equations for ionization dynamics in the crystal}
\label{sec:4}

In order to solve the evolution problem for the electron density in the crystal it is convenient to separate the whole system into three subsystems: the bound electrons (discrete spectrum), the free electron gas (continuous spectrum), and the electromagnetic field.

1) It has been shown in many papers (for example, \cite{Santra-Son} and citations therein) that the most efficient way to describe the dynamics  of the bound electrons is obtained by studying the time dependence of any electron configuration of the atom. Since a set of bound electrons at a given time represents a certain atomic configuration, their evolution can be described as time-dependent changes between different possible configurations. It may start from the neutral atom and may finish with a fully ionized atom. If one writes $P_\lambda (t)$ for the probability of the $\lambda$ configuration at an arbitrary moment of time, then the initial condition for this function corresponds to the case where all atoms are in the ground (neutral) state:

\begin{eqnarray}
\label{1b}
    P_\lambda (0) = \delta_{\lambda, 0}.
\end{eqnarray}

One should also stress the normalization condition for the whole set of atomic configuration probabilities that should be fulfilled for any arbitrary moment of time:

\begin{eqnarray}
\label{1a}
    \sum_\lambda P_\lambda (t) = 1.
\end{eqnarray}

With this definition, the   population  of the atomic level  $Q_l(t) = \left\langle |a_l(t)|^2 \right\rangle$ in the scattering factor (\ref{12a}) averaged over all configurations is defined as follows:

\begin{eqnarray}
\label{11a}
    F(\vec h, t) = \sum_{l \leq l_m,j} Q_l(t) F_{l,j}(\vec{h})e^{i \vec h \vec R_j}, \nonumber \\
    Q_l(t) =    \sum_\lambda P_\lambda (t)g_{l,\lambda}|a_l(t)|^2,
\end{eqnarray}
\noindent where $g_{l,\lambda}$ is the degeneracy of this level in the configuration $\lambda$.

2) Electrons of the continuous spectrum appear due to photoionization, Auger recombination, and electron-impact ionization. This subsystem can be described in terms of a classical one-particle distribution function $f(\vec r, \vec p , t)$ normalized as follows:

\begin{eqnarray}
\label{1c}
    \quad \int f(\vec r, \vec p , t) d \vec p d \vec r = n_e (t),
\end{eqnarray}

\noindent where $n_e (t)$ is the total number of free electrons per unit cell.

This subsystem includes all excited electrons as well because for any excitation they occupy the conduction bands following the free electron approximation for overlapping electron shells of atoms in a crystal.

At the initial moment of time there are no free electrons, which corresponds to the following condition:

\begin{eqnarray}
    f(\vec r, \vec p , 0) = 0.
\end{eqnarray}

One should also note that although the photoionization cross section is not isotropic over the ejected electron direction \cite{LandauV4} , the multiple electron--electron collisions lead to the loss of information about the initial velocity directions, so that the distribution function $f(p)$ can be assumed to be isotropic over the momentum variable \cite{LandauV10}.

3) The electromagnetic field is described by the wave packet

\begin{eqnarray}
\label{1d}
    \vec A (\vec r, t) = \vec e_s \Phi(\vec r,t) e^{i (\vec k \vec r - \omega t)}, \nonumber\\
    I(\vec r, t) =   |\Phi(\vec r,t)|^2,
\end{eqnarray}

\noindent where $I(\vec r, t)$ is the intensity distribution function. Using the kinematical approximation of X-ray diffraction, the evolution of the electromagnetic field is not taken into account.

Let us consider the general form of the rate equations describing the atomic population dynamics \cite{Santra-Son}:

\begin{eqnarray}
\label{3_1}  \frac{dP_\lambda}{dt} = \sum_{\mu \neq \lambda} (W_{\mu \lambda} P_{\mu} - W_{\lambda \mu} P_{\lambda}),
\end{eqnarray}

\noindent where $P_{\lambda}$ is the probability of the system's occupying a configuration with index $\lambda$ and $W_{\lambda \mu}$ is the probability of a  transition between the configuration $\lambda$ to $\mu$ in unit time.

Transitions between various atomic configurations during the XFEL pulse propagation are mainly caused by photoionization, Auger decay, electron-impact ionization, and three-body recombination. The photoionization rate is given by

\begin{eqnarray}
    W^{(Ph)}_{\lambda \mu} (t) = \sigma^{(Ph)}_{\lambda \mu} J(t),
\end{eqnarray}

\noindent where $\sigma^{(Ph)}_{\lambda \mu}$ is the cross-section of the photoionization process that corresponds to the transition from configuration $\lambda$ to $\mu$ and $J(t)$ is the photon flux function.

For the time-independent Auger process rate $W^{(Ag)}_{\lambda \mu}$ we use the expressions given in \cite{Santra-Son, SantraPhoto} and modify them with all ionization potentials calculated in the framework of ECM \cite{FerTriguk}.

The electron-impact ionization rate can be deduced on the basis of a collision integral calculation and has the following explicit form:

\begin{eqnarray}
\label{2a}
    W^{(eii)}_{\lambda \mu} = \frac{n_a}{2} \int v^3 f(v) dv \int \frac{d \sigma_{\lambda \mu}^{(eii)} (v'|v)}{dv'} dv',
\end{eqnarray}

\noindent where $n_a$ is the number of atoms per unit cell and the parameter dependence of the cross-section is organized in the way $(v_{fin}|v_{ini})$.

Using the principle of detailed balance \cite{Hau-Riege-book, LandauV10}, the rate of the three-body recombination process can be deduced on the basis of the electron-impact ionization rate:

\begin{eqnarray}
\label{2b}
    W^{(tbr)}_{\lambda \mu} = \left( \frac{2 \pi \hbar}{m_e} \right)^3 \frac{n^2_a}{2} \int f(v) dv \int {v'}^3 \frac{ d \sigma_{\mu \lambda}^{(eii)} (v|v')}{dv} f(\sqrt{{v'}^2-v^2-\frac{2}{m_e} E_{\mu \lambda}}) dv',
\end{eqnarray}

\noindent where $E_{\mu \lambda}$ is the ionization potential that corresponds to the transition from configuration $\mu$ to $\lambda$.

It is important to stress that as long as the rates (\ref{2a})--(\ref{2b}) depend on the electron density function (see below), the subsystems of free and bound electrons are coupled.

The dynamics of the free electron gas density function is described by the Boltzmann kinetic equation and has the form \cite{LandauV10}

\begin{eqnarray}
\label{10}  \frac{d f(\vec r, \vec p, t) }{dt} = \frac{\partial f(\vec r, \vec p, t) }{\partial t} + \vec v \vec {\nabla}_{\vec r} f(\vec r, \vec p, t)  + \vec F \vec {\nabla}_{\vec p} f(\vec r, \vec p, t)  = I_B[ f(\vec r, \vec p, t)].
\end{eqnarray}

For simplicity and insight into the ongoing processes, let us make a number of additional assumptions. First of all, let us suppose that the system remains homogeneous in the lateral direction during the field--matter interaction due to the fact that the beam size in this direction is much larger than the size of a crystal cell. This means that all functions depend only on $z$ (the axis parallel to the wave vector) and $t$; the wave front itself depends on the variable $z-ct$.

Furthermore, the only vector that could cause an anisotropy in momentum space is the photon momentum, so that the anisotropy parameter

\begin{eqnarray}
    \xi_a \approx \frac{k_{ph}}{p_e} \sim \sqrt{ \frac{\hbar \omega }{m_e c^2}} \sim 0.14 \ll 1,
\end{eqnarray}

\noindent and due to thermalization the density function can be considered approximately isotropic over the momentum directions.

In the non-relativistic case, the net force $\vec F$ acting on an electron is defined by the uncompensated Coulomb field created by the other electrons of the continuous spectrum and the ionized atoms. This force becomes essential if the photon pulse has left the crystal but can be neglected during the passage of the pulse through the crystal. Moreover, in the non-relativistic case with the assumptions mentioned above, the diffusion term yields

\begin{eqnarray}
    \vec v \vec {\nabla}_{\vec r} f(\vec r, \vec p, t) \sim \frac {v}{c} \frac{\partial f }{\partial t} \ll \frac{\partial f }{\partial t}
\end{eqnarray}

\noindent and can be neglected as well.

As a result of these approximations, one can reduce the initial Boltzmann equation (\ref{10}) to the form

\begin{eqnarray}
\label{10red}
    \frac{\partial f(v, t) }{\partial t} = I_B[ f ].
\end{eqnarray}

In the collision integral  $I_B$, the following transitions should be taken into account: 1) electron-impact ionization of atoms (ions); 2) three-body recombination; 3) electron--electron elastic scattering. The corresponding collision integrals can be written as

\begin{eqnarray}
\label{11a}
    I^{(eii)}_B = n_a \left[ \frac{1}{v^2} \sum_{\lambda, \mu} P_{\lambda} \int {v'}^3 \frac{d \sigma_{\lambda \mu}^{(eii)} (v|v')}{dv} f(v') dv' \right. \notag\\
    -\left. \frac{1}{2} v f(v) \sum_{\lambda, \mu} P_{\lambda} \int \frac{ d \sigma_{\lambda \mu}^{(eii)} (v'|v)}{dv'}dv' \right].
\end{eqnarray}

In order to derive the three-body recombination collision integral one can use the principle of detailed balance \cite{LandauV10}, so that the corresponding cross-section can be obtained on the basis of the electron-impact ionization cross-section:

\begin{eqnarray}
\label{11b}
    I^{(tbr)}_B = \left( \frac{2 \pi \hbar}{m_e} \right)^3 n^2_a \left[ \frac{1}{2} v \sum_{\lambda, \mu} P_{\lambda} \int \frac{ d \sigma_{\mu \lambda}^{(eii)} (v'|v)}{dv'} f(v') f(\sqrt{v^2-{v'}^2-\frac{2}{m_e} E_{\mu \lambda}}) dv' \right.
    \notag\\
    - \left. \frac{f(v)}{v^2} \sum_{\lambda, \mu} P_\lambda \int {v'}^3 \frac{ d \sigma_{\mu \lambda}^{(eii)} (v|v')}{dv} f(\sqrt{{v'}^2-v^2-\frac{2}{m_e} E_{\mu \lambda}}) dv' \right].
\end{eqnarray}

In order to describe the elastic electron--electron scattering we implement the scheme of relaxation dynamics for particle systems with Coulomb interaction as introduced by \cite{MacRos}.

It is important to stress again that due to the dependence of the collision integrals (\ref{11a})--(\ref{11b}) on the atomic configuration the probabilities $P_\lambda$ shown in Equations (\ref{3_1}) and (\ref{10red}) are coupled and must be  solved simultaneously. However, as long as we use ECM \cite{FerTriguk}, all cross-sections  introduced in the system of master equations can be calculated analytically (see the Appendix) with the necessary accuracy. The latter makes numerical simulations less expensive in time and resources.

\section{Numerical results for atomic populations}
\label{sec:5}

In order to simulate the population dynamics we implemented the algorithm \cite{ELENDIF} to solve the Boltzmann equation and the system of rate equations. The XFEL pulse used for calculations was specified to have a photon energy of 8 keV, a photon number of $10^{12}$ per pulse, beam size of 1 $\mu$m$^2$ (thus the fluence being $1.6 $\ mJ/$\mu$m$^2$), Gaussian shape with full duration of 40 fs (17 fs FWHM). All calculations have been done for the example of a Silicon crystal.

The energy of the Si K-line (1.8 keV) is more than four times smaller than the photon energy of 8 keV, resulting in a non-resonant photon to atom interaction. In order to estimate the electron density evolution for a photon energy closer to the Silicon K-edge, where non-resonant effects become non-negligible, we performed additional simulation for a 3 keV pulse with the same characteristics as defined above (the fluence being 0.6 mJ/$\mu$m$^2$ in this case).

\begin{figure}[tb]
\includegraphics[scale=0.70]{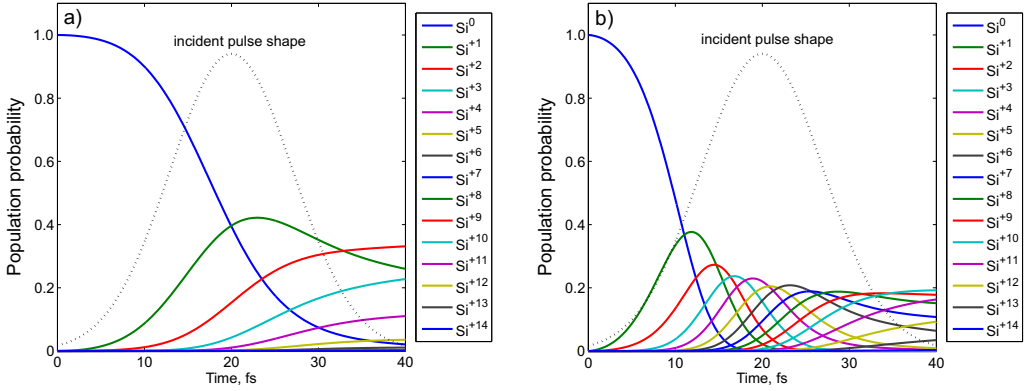}
\caption{Atomic population probabilities for Si crystal as the function of time: (a) -- 8 keV pulse, (b) -- 3 keV pulse.}
\label{picnum1}
\end{figure}

\begin{figure}[tb]
\includegraphics[scale=0.70]{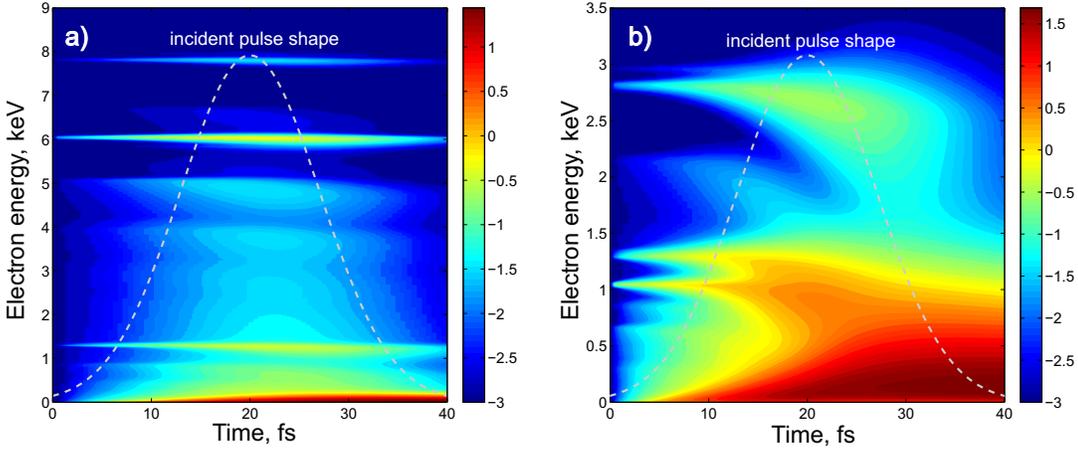}
\caption{Free electron density as the function of time and energy:  (a) -- 8 keV pulse, (b) -- 3 keV pulse.}
\label{picnum3}
\end{figure}

\begin{figure}[tb]
\includegraphics[scale=0.70]{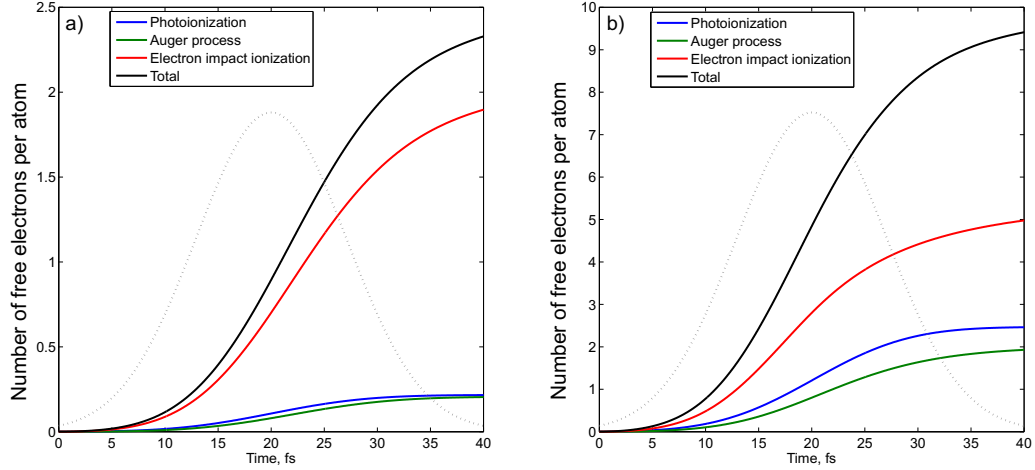}
\caption{Contribution of different channels and total yield of free electrons per atom:  (a) -- 8 keV pulse, (b) -- 3 keV pulse.}
\label{picnum2}
\end{figure}

\begin{figure}[tb]
\includegraphics[scale=0.70]{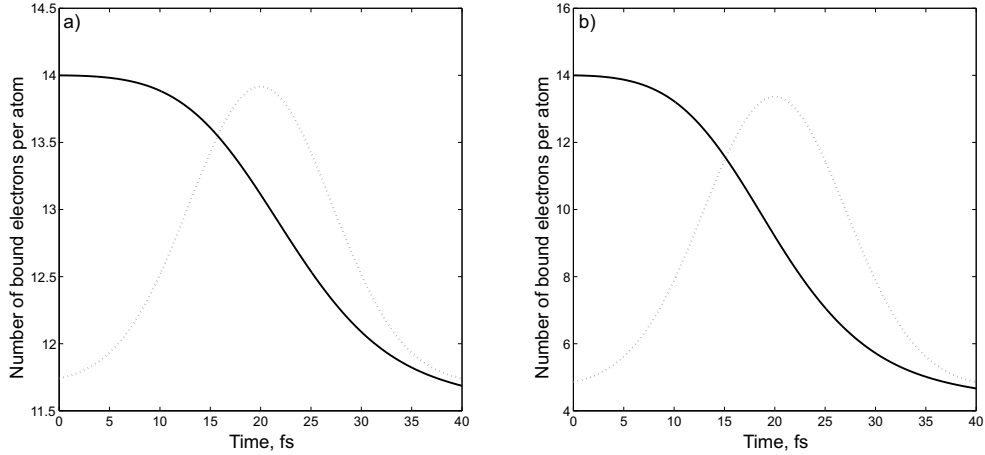}
\caption{Average number of bound electrons per atom as the function of time:  (a) -- 8 keV pulse, (b) -- 3 keV pulse.}
\label{picnum4}
\end{figure}

\begin{figure}[tb]
\includegraphics[scale=0.60]{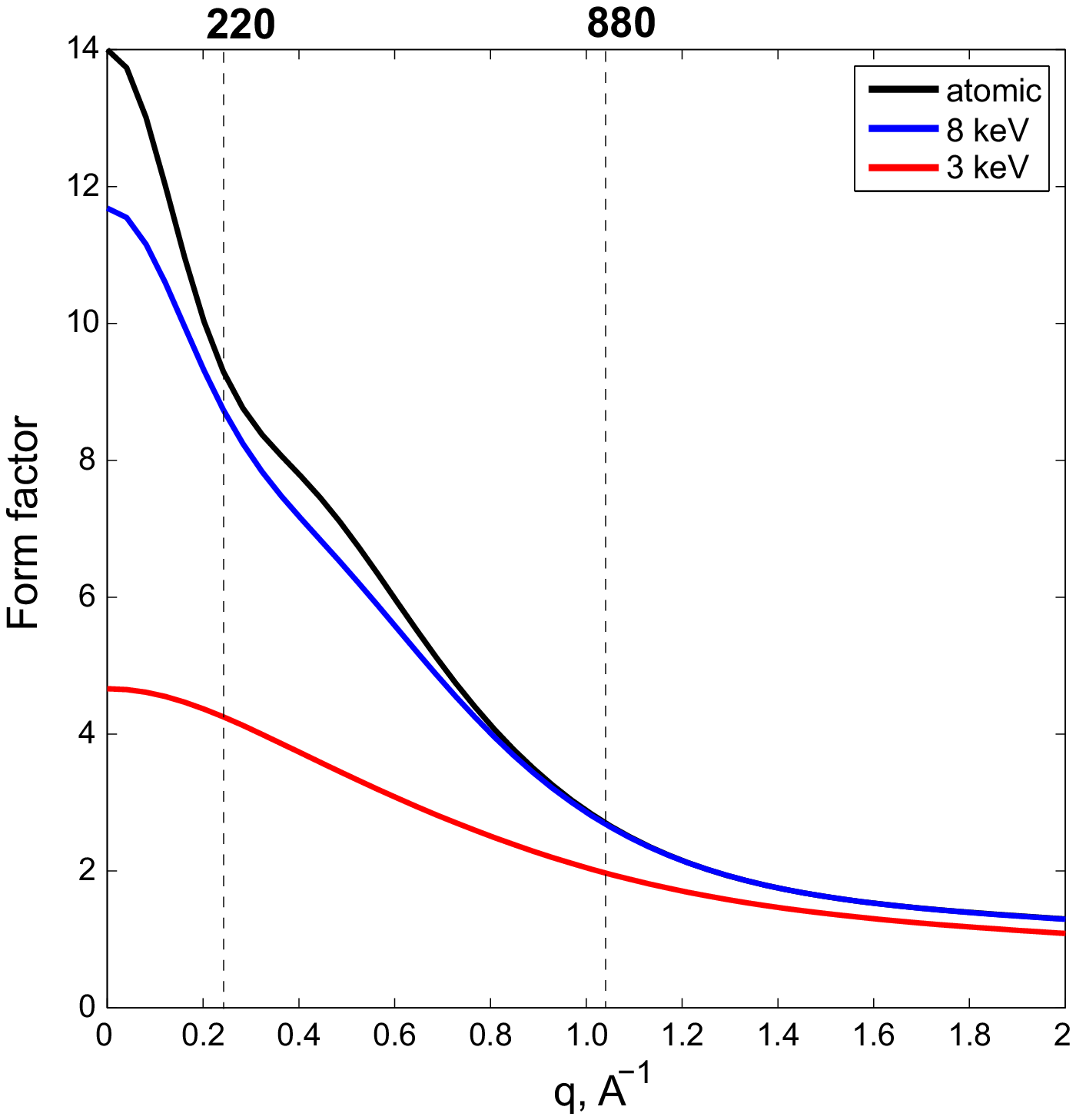}
\caption{ASF as a function of $q$ for the conventional case and after the passing of 8 kev and 3 keV pulses.}
\label{Fig.14}
\end{figure}

Fig. \ref{picnum1} shows the probability of finding differently ionized ions in the Silicon crystal as a function of time. It shows that the number of neutral atoms decreases during the time of interaction between the photon pulse and the crystal. At 8 keV photon energy the population probability decreases almost to zero by the end of the pulse and for 3 keV case it decreases completely to zero already at half of the pulse length. The latter is remarkable considering the fact that the pulse energy is about 1.2 keV above the threshold of atomic K-shell ionization. Here most of the populated states are $+9$ and $+10$ at the end of the pulse. In the non-resonant case at 8 keV the interaction between the XFEL pulse and the electron subsystem of the atom is weak, so that the atoms are not so deeply ionized and the mostly populated states are ions with $+1$, $+2$ and $+3$ ionization charges.

Fig. \ref{picnum3} shows the distribution of kinetic energy of the free electrons as a function of time. At 8 keV, i.e., in the non-resonant case, one can see three vivid energy bands varying in time: the top band (at about 6 keV) describes the energy of the photo electrons, the middle band (about 1.3 keV) corresponds to the energy of the Auger electrons, and the range close to zero energy describes the secondary electrons that appear due to the electron-impact ionization process. In contrast to this, the 3 keV result shows a broad spectrum corresponding to the photo (both spikes at about 1.0 keV and 2.8 keV) and Auger (middle spike at about 1.3 keV) electrons. Additionally the bands are broaden due to the fact that every step of ionization is accompanied by a certain decrease of the ionization potential and subsequently a reduction of the energy of every successive photo electron. Moreover, the free electrons undergo elastic and inelastic scattering, which also results in a broadening of the energy distribution.

Fig. \ref{picnum2} shows the total number of free electrons per atom in the crystal unit cell and the contribution of the different ionization channels in time. One can conclude that in both cases, the near-resonant and the non-resonant one, the electron-impact ionization channel plays the dominant role for the creation of free electrons. The respective yield of free electrons via this process for 8 keV case is almost seven times higher than those of photoionization and Auger processes. For 3 keV case the relative contribution of the electron-impact ionization channel is about two times larger than that of the photoionization and Auger recombination but in absolute numbers two times larger than for 8 keV photons.

\section{Evolution of the atomic scattering factor}
\label{sec:6}

\begin{figure}[htb]
\includegraphics[scale=0.6]{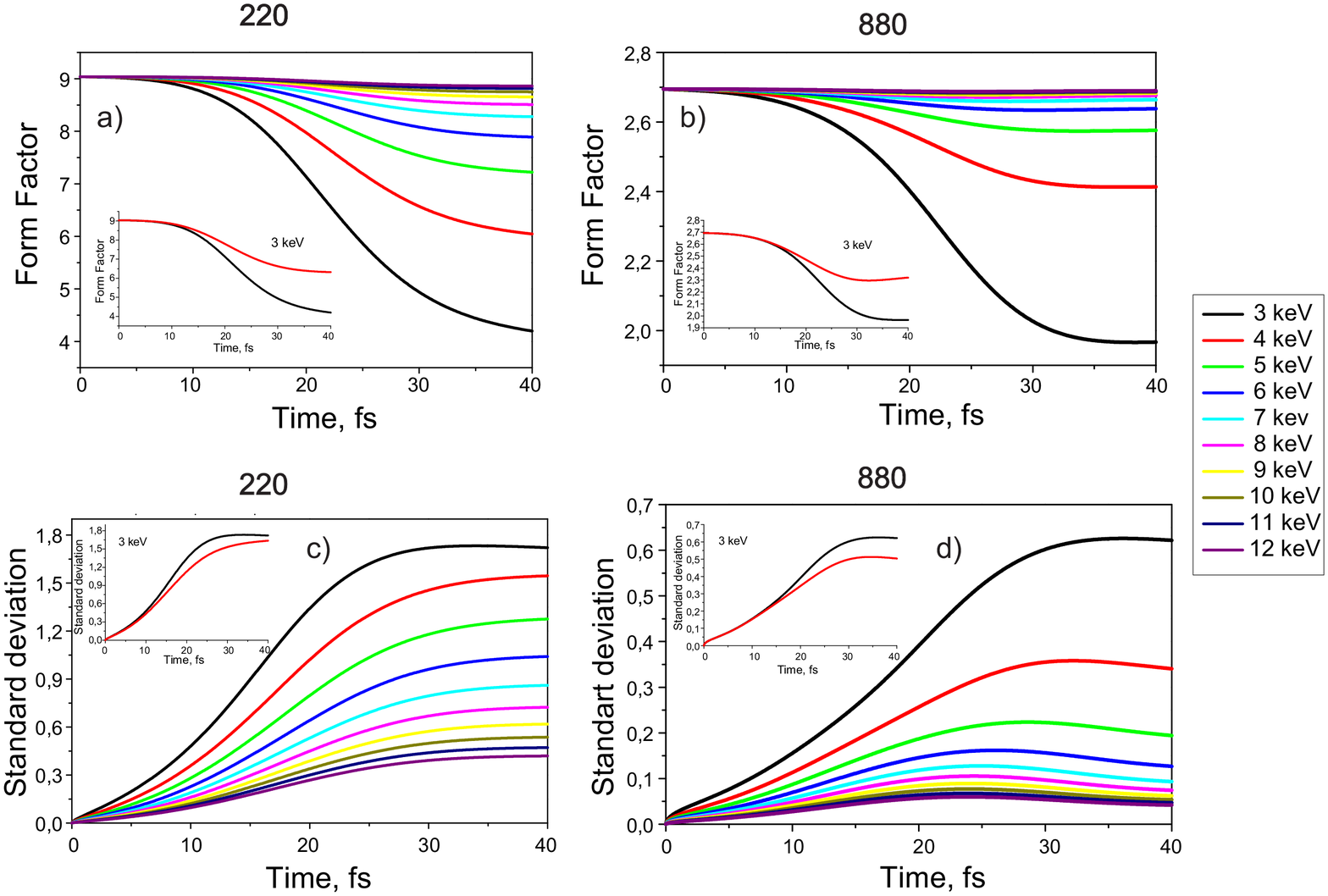}
\caption{Evolution of the average ASF $\bar{F}(\vec h,t)$ (a, b)and its standard deviation $\Delta F(\vec h,t)$ (c, d) as functions of the photon energy. Inset: Evolution of the average ASF $\bar{F}(\vec h,t)$ and its standard deviation $\Delta F(\vec h,t)$ without (red line) and including (black line) the contribution of free electrons for the 3 keV case. (220) and (880) reflections are considered }
\label{Fig.10}
\end{figure}

\begin{figure}[htb]
\includegraphics[scale=0.6]{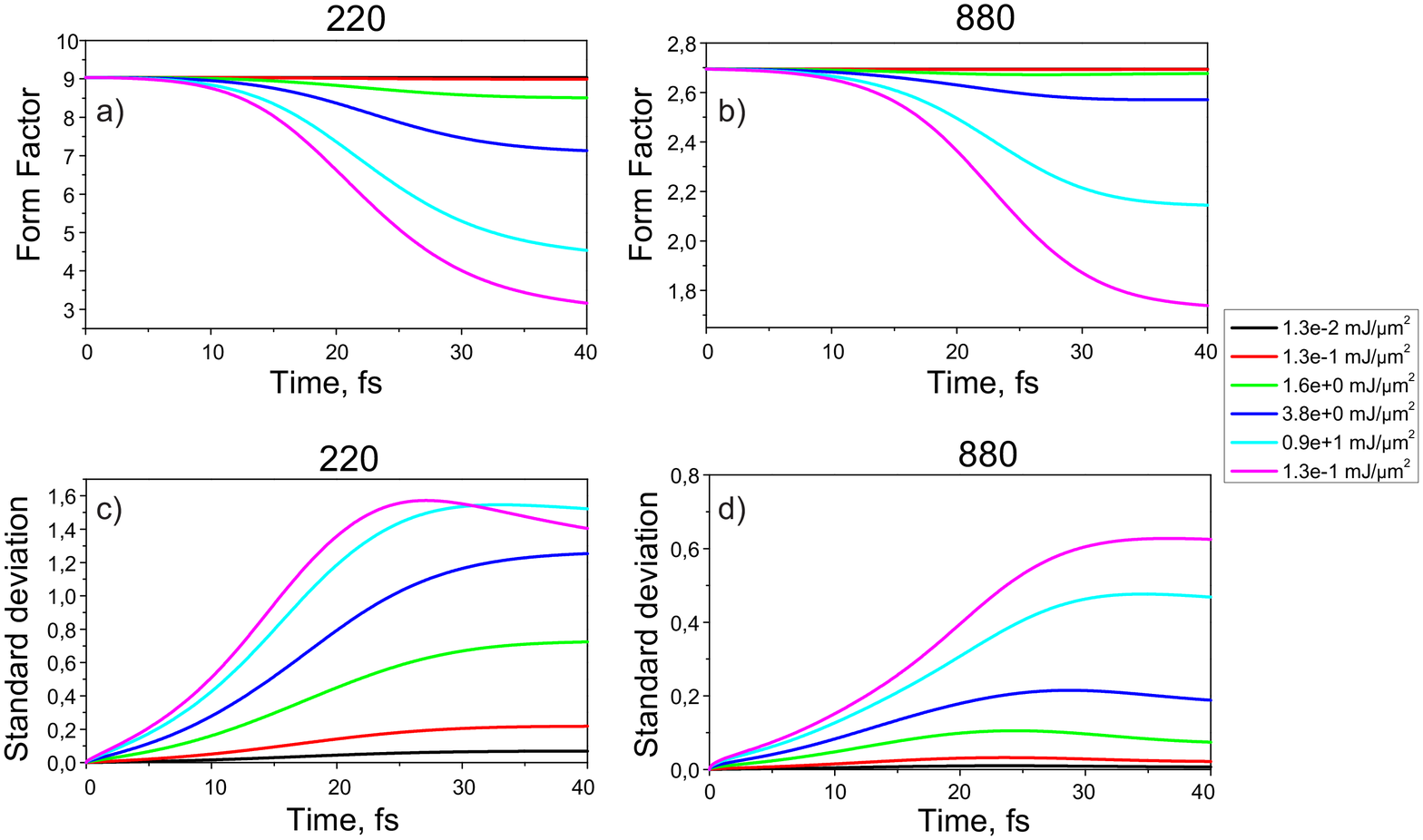}
\caption{Dependence of  the average ASF (a, b) and its standard deviation (c, d) on the fluence of the XFEL pulse}
\label{Fig.11}
\end{figure}

The most relevant quantity for the formation of the diffraction peak is the average value ASF $\bar{F}(\vec q,t)$ describing the number of scattering electrons as a function of the momentum transfer $q = \frac{\sin \theta}{\lambda}$, where $\theta$ is the scattering angle and $\lambda$ is the photon wave length. The statistical character of the ionization processes means that the ASF at a moment of time $t$ is a random value which depends on the probabilities of finding a certain electron configuration of the atom $P_{\lambda}(t)$. Let us define the amount of the average ASF $\bar{F}(\vec q,t)$ and its standard deviation $\Delta F(\vec q,t)$ as follows:

\begin{eqnarray}
\label{23}
    \bar{F}(\vec q,t) = \sum_{\lambda} F_{\lambda}(\vec q) P_{\lambda}(t), \nonumber\\
    \Delta F(\vec q,t) = \sqrt{\overline{F^2}(\vec q,t) - \bar{F}^2(\vec q,t) }, \\
    \overline{F^2}(\vec q,t) =  \sum_{\lambda} F^2_{\lambda}(\vec q) P_{\lambda}(t),\nonumber
\end{eqnarray}

\noindent where $F_{\lambda}(\vec q)$ is the stationary ASF value for the atomic configuration ${\lambda}$ at the momentum transfer $\vec q$. Since the anomalous dispersion term is omitted  we do not consider the energy range close to the exact resonance energy.

The calculation of the ASF value with probabilities $P_{\lambda}(t)$ related to one cell is performed by use of the ergodic hypothesis \cite{LandauV10} for the statistical ensemble of the atoms in the whole crystal. It is also supposed that the fluctuations of the ASF for atoms in different cells are not correlated. In this case the ASF dispersion contributes only to the X-ray diffuse scattering background and does not change the intensity of the coherent diffraction peak \cite{Vartanyants}.

Because the free electron distribution is broad in real space the value of the ASF   mainly depends on the number of bound electrons in the atoms/ions. Fig. \ref{picnum4} shows the alteration of this number during the pulse length. It becomes evident that atoms lose about 9 bound electrons in the near-resonant case, whereas in the non-resonant case the drop is less than 3 electrons per atom.

Fig. \ref{Fig.14} shows the ASF values as a function of momentum transfer, $q$,  for three cases: the neutral free atom and the time-averaged ASF after the passage of the 3 keV and 8 keV XFEL pulses, respectively. We find a significant reduction in the ASF in the near-resonant ($3 \ keV$) case over all values of $q$. On the other hand, the drop in the ASF is small in the non-resonant case ($8 \ keV$) and is substantial at low $q$ values only.

The properties of the ASF are studied in more detail for two different diffraction peaks: the 220 Bragg peak at $q=0.26 {\AA}^{-1}$ and the 880 Bragg peak at $q= 1.04 {\AA}^{-1}$ (see the dotted line in Fig. \ref{Fig.14}) where the first one is affected by changes in both the valence and core shells but the second one mainly depends on the core density only.

Fig. \ref{Fig.10} shows the time evolution of  the  ASF at the $q$ position of the 220 and 880 Bragg reflections as a function of photon energy between 3 keV and 12 keV. Without interaction with the XFEL pulse the  ASF is about 9.0 for the 220 but 2.7 for the 880 Bragg peak for all photon energies. Both values drop during the time of interaction of the XFEL pulse with the crystal. The amount of this drop increases with decreasing energy difference to the K-absorption edge. At 3 keV and for the 220 reflection the total ASF decreases by $50\%$ during the XFEL pulse of 40 fs, for 880 this drop is 35\%. The inset of Fig. \ref{Fig.10} shows the drop for the 3 keV pulse case with and without the contribution of the free electrons. It becomes evident that the free electrons contribute by about $20\%$ and $13\%$ to the time-dependent drop of the 220 and 880 ASF, respectively. At the same time, the 8 keV pulse causes less photoionization damage, so that the ASF drop is only $7\%$ for the 220 and $2\%$ for the 880 Bragg reflections in the non-resonant case.

The critical point of X-ray diffraction with XFEL pulses is to find the photon intensity that initiates complete ionization of the atom during a time faster than that necessary for the formation of the diffraction peak, i.e., faster than the pulse time. This threshold intensity can be determined using the numerical results shown in Fig. \ref{Fig.11}. It shows the flux dependence of both reflections and their standard deviation. It becomes evident that the form factor drop is dramatic if the fluence exceeds 1.6 mJ/$\mu$m$^2$).

In the framework of the kinematical theory, the diffraction peak intensity is defined by the square of   the ASF   from all atoms  \cite{Landau8}. In the case of XFEL pulse diffraction, it is the fluctuating value that should be averaged for all configurations:

\begin{eqnarray}
\label{24}
\bar{R}(\vec q,t) \sim \left\langle  \sum_a \sum_b F_a (\vec q)F_b^* (\vec q) e^{i \vec q (\vec R_a - \vec R_b)}  \right\rangle,
\end{eqnarray}
\noindent where the symbol $<...>$ means the average over all configurations defined in the formula (\ref{11a}) and the summation is performed over the coordinates $\vec R_a, \vec R_b$ of the same atoms with ASF $F_a (\vec q)$  in all unit cells of the crystal.

It was mentioned above that the average ASF is supposed to be the same for all unit cells and its fluctuations are not correlated. This allows one to use the formula

\begin{eqnarray}
\label{24a}
  \left\langle    F_a (\vec q)F_b^* (\vec q)  \right\rangle  =  \bar{F}  (\vec q,t )\bar{F}^* (\vec q,t) + \Delta^2 F(\vec q ,t) \delta_{ab}; \nonumber\\
\bar{R}(\vec q,t) \sim   |\bar{F}  (\vec q,t )|^2 |\sum_a   e^{i \vec q  \vec R_a  }|^2 + N \Delta^2 F(\vec q ,t),
\end{eqnarray}
\noindent where $N$ is the total number of unit cells in the crystal.

The first term in (\ref{24a}) defines the coherent diffraction intensity in accordance with the identity \cite{solid}

\begin{eqnarray}
\label{24b}
\bar{R}(\vec q,t)\equiv \sum_{\vec h}\bar{R}(\vec h,t)\delta_{\vec q, \vec h}   \sim   N^2 \sum_{\vec h} |\bar{F}  (\vec h,t )|^2  \delta_{\vec q, \vec h}.
\end{eqnarray}

This value is proportional to $N^2$ and is significantly larger than the diffuse scattering background defined by the ASF fluctuations in the second term in (\ref{24a}).

Compared to the intensity of conventional diffraction, the change of diffraction intensity induced by an XFEL pulse $R_0(\vec h)$ can be characterized by the value

\begin{eqnarray}
\label{25}
    \frac{R}{R_0} (N_{ph}, \omega)  = \frac{1}{ R_0(\vec h) }\int_{-\infty}^{\infty} I(t) \bar{R}(\vec h,t) dt,
\end{eqnarray}

\noindent that is a function of the number of photons in the pulse $N_{ph}$ (photon flux) and their frequency $\omega$; the intensity slope function $I(t)$ is defined in (\ref{1d}); $R_0(\vec h) \equiv R(\vec h,-\infty)$.

As was shown above, the standard deviation of a Bragg reflection is significantly less than the average ASF, so the expression (\ref{25}) can be written as follows

\begin{eqnarray}
\label{26}
    \frac{R}{R_0} (N_{ph}, \omega)  \approx  \frac{1}{ |F_0(\vec h)|^2 }\int_{-\infty}^{\infty} I(t) |\bar{F}(\vec h,t)|^2 dt.
\end{eqnarray}

\begin{figure}[tb]
\includegraphics[scale=0.35]{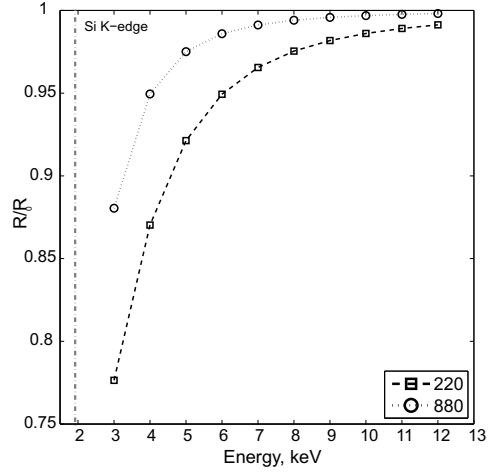}
\caption{Integral intensity of the XFEL pulse diffraction compared with the conventional low energy diffraction as a function of photon energy with $10^{12}$ photons per pulse.}
\label{Fig.12.energy}
\end{figure}

\begin{figure}[tb]
\includegraphics[scale=0.35]{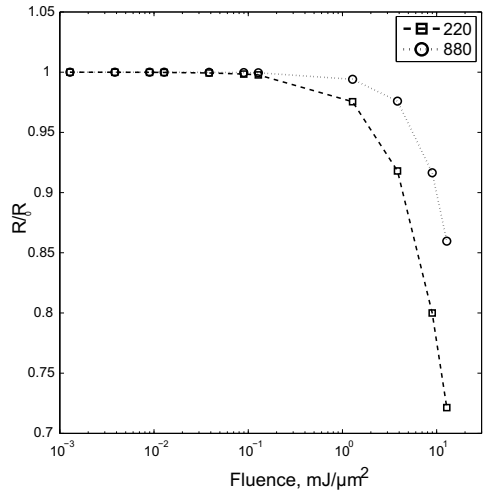}
\caption{Integral intensity of the XFEL pulse diffraction compared with the conventional low energy diffraction as a function of fluence with 8 keV photon energy.}
\label{Fig.13.flux}
\end{figure}

Fig. \ref{Fig.12.energy} and Fig. \ref{Fig.13.flux} show the dependence of this ratio as functions of photon energy and fluence. Fig. \ref{Fig.12.energy} demonstrates that the diffraction intensity decreases in a non-linear manner if the photon energy approaches the Si K-edge. The deviation from unity  is less than $5\%$ for 8 keV and reaches about $20\%$ at 3 keV and will decrease further for energies closer to the K-edge energy. The calculation of the fluence dependence of $\frac{R}{R_0}$ at 8 keV demonstrates a dramatic drop if the fluence exceeds $1.6 $\ mJ/$\mu$m$^2$ (see above).

\section{Discussion and conclusions}
\label{sec:7}

A numerical algorithm and software were developed for calculation of the X-ray diffraction intensity during the propagation of an intense XFEL fs-pulse through a crystal. Together with photoionization and Auger processes  we considered additional processes related to the free electrons generated in the conduction band of the solid state.

\begin{figure}[tb]
\includegraphics[scale=0.75]{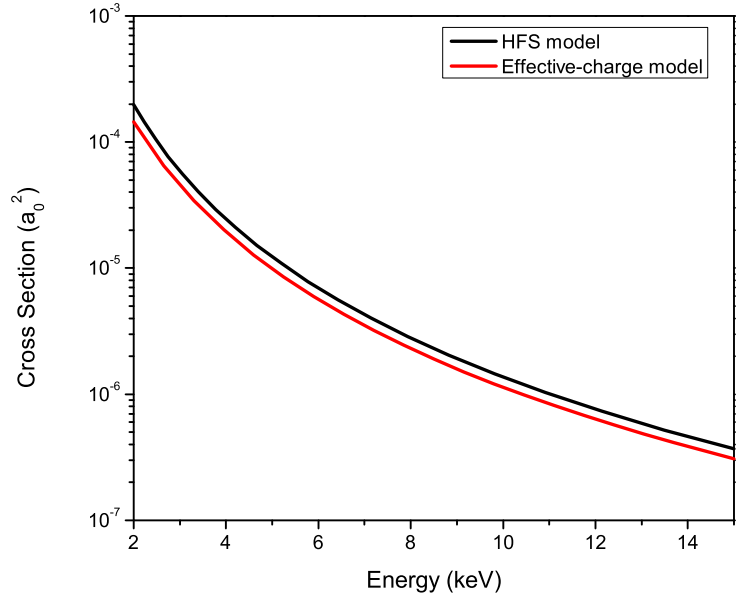}
\caption{Photoionization cross-section for 1s-shell of neutral carbon calculated within HFS approach (black line) and ECM (red line).}
\label{comp:1}
\end{figure}

\begin{figure}[htb]
\includegraphics[scale=0.65]{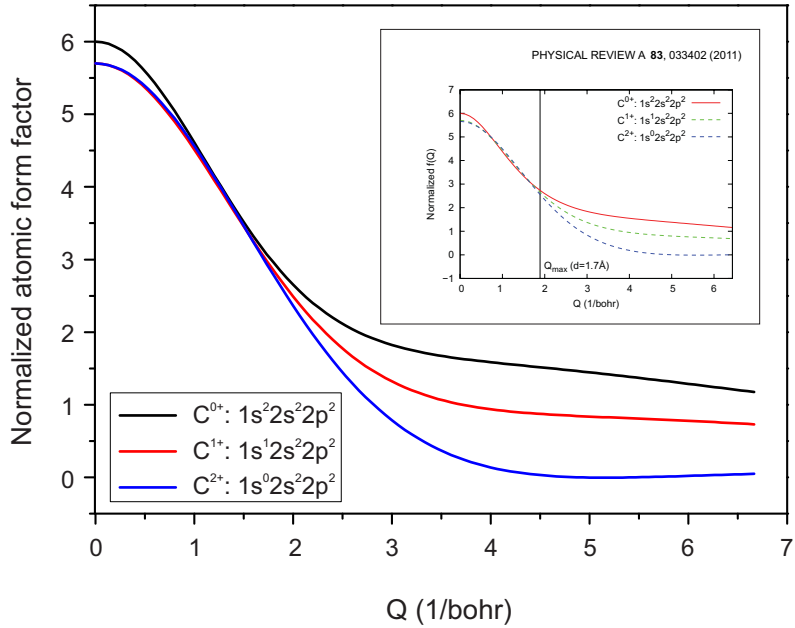}
\caption{Normalized (in accordance with \cite{Santra-Son}) atomic scattering factor for neutral (black line), single core-hole (red line) and double core-hole (blue line) states of carbon (the agreement with \cite{Santra-Son} is very good).}
\label{comp:2}
\end{figure}

\begin{figure}[htb]
\includegraphics[scale=0.65]{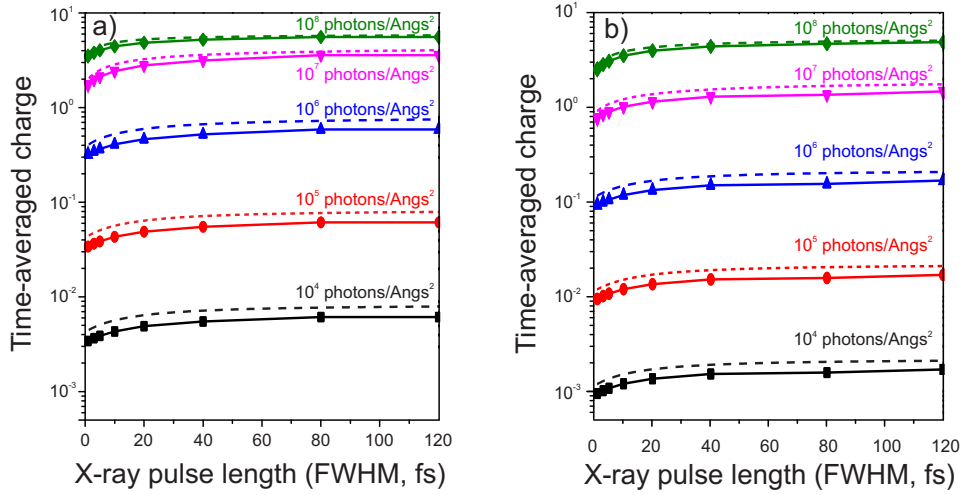}
\caption{Time-averaged charge as a function of pulse duration: (a) -- 8 keV pulse; (b) -- 12 keV pulse. Solid lines correspond to CEIX calculation, dashed lines to data from \cite{Santra-Son}.}
\label{comp:3}
\end{figure}

\begin{figure}[!htb]
\includegraphics[scale=0.65]{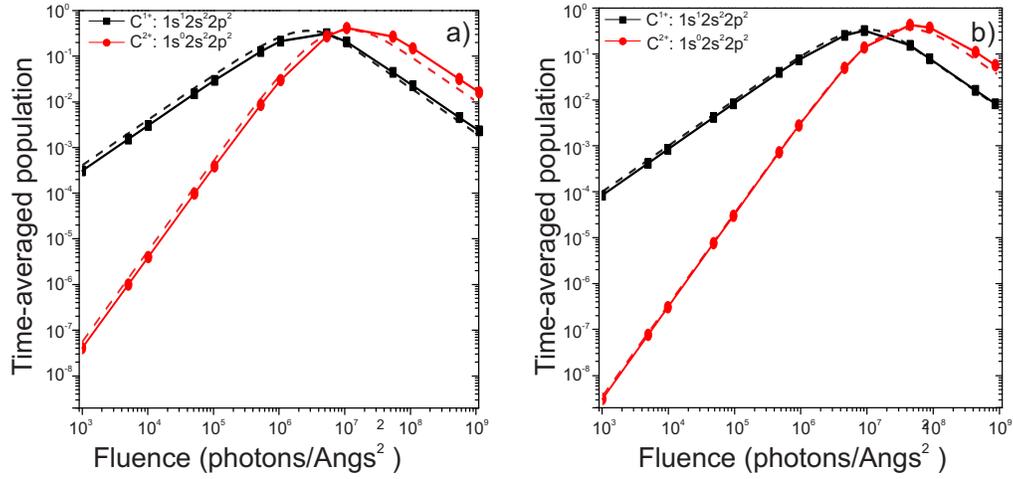}
\caption{Time-averaged atomic population probabilities of the single core-hole and double core-hole states of carbon: (a) -- 8 keV pulse; (b) -- 12 keV pulse. Solid lines correspond to CEIX calculation, dashed lines to data from \cite{Santra-Son}. }
\label{comp:4}
\end{figure}

In order to check both the validity of the results predicted by  ECM  and the stability of the numerical algorithm for solving the system of master equations, we simulated the atomic dynamics of carbon. This system has been calculated by Santra et al \cite{Santra-Son} using a full numerical treatment in terms of the Hartree--Fock--Slater (HFS) model \cite{LANL}. As a first test, we compared the photoionization cross-sections predicted by ECM with the results obtained by the HFS model. The comparison is shown in Fig. \ref{comp:1} -- both have the same funcional behavior. The slight shift between the ECM and HFS values occurs due to the fact that for reasons of simplicity, we made a rough estimation for the continuous spectrum radial wave function.

Calculations of the normalized ASF for certain atomic configurations (neutral, single core-hole and double core-hole states of carbon) by ECM are shown in Fig. \ref{comp:2}. One can conclude that these quantities are in very good agreement with the results shown in Fig. 1 of \cite{Santra-Son} over the whole range of the momentum transfer values.

Fig. \ref{comp:3} and Fig. \ref{comp:4} show the results of the simulation of the electron dynamics in carbon for the cases of pulses of 8 keV and 12 keV. Our data are in good agreement with the results shown in Fig. 2 and Fig. 3 of \cite{Santra-Son}. However, one should note that there is a small shift between the corresponding extrema of atomic population probabilities (see Fig. \ref{comp:4} and Fig. 3 of \cite{Santra-Son}) caused mainly by the fact that we implemented smaller values for the photoionization cross-sections for the reason mentioned above.

The good agreement with the results of HFS for carbon justify the validity of our code for the case of Silicon. The results of the present paper lead to the following general conclusions:

According to Fig. \ref{picnum2}, the role of the free electrons is dominant via the process of electron impact ionization.

According to Fig. \ref{Fig.12.energy}, our approach remains valid for photon energies about 1 keV above the K-edge. However, in order to make the simulation more precise and avoid additional errors, the accuracy of the photoionization cross sections within the ECM should be improved for this energy region. For Silicon this may happen at a photon energy of 2.5 keV. Further decrease of the photon energy will result in a decrease of the ionization potential up to the value where the ionization potential becomes deeper than the photon energy itself, where single photon transitions from the K-shell become forbidden. Generally, the approach to the solution of the rate equations becomes invalid in close vicinity to the exact resonance. Here, one should use the density matrix method in order to take into account both diagonal and non-diagonal elements for the solution of the evolution problem. A more exact treatment in terms of quantum mechanics is needed in order to consider quantum coherence effects (Rabi oscillation) that are expected if the photon energy exactly matches the energy of transition.  The coherence effects become significant only if Eq. (\ref{8aY}) is non-local in time, i.e., if $Y(t-t')$ has a significant time spread in comparison with the pulse duration, or, turning to the frequency domain, if $Y(\omega)$ is sharp in comparison with the spectral width of the pulse slope function. In this case the resonance can take place and the system of Eqs. (\ref{9Y}) and (\ref{11rabi}) should be used to calculate the amplitudes.
However, if the frequency of the X-ray pulse corresponds to the transition to the continuous part of the spectrum, $Y(\omega)$ covers a wide range of the X-ray frequency that is  broader than the spectral width of the pulse slope function. Then non-Markovian effects can be neglected, and we come to the rate equations in the form (\ref{10rateeq}) for the occupation probabilities.

The general result of our numerical investigation consists in the predicted time dependence of the atomic form factor.
As shown in Fig. \ref{Fig.10}, the ASF decreases during the propagation of the intense XFEL pulse through the crystal. This results in a drop in the diffraction intensity during the pulse propagation.
Due to photoionization and other processes, the amount of this drop at the end of the pulse increases if the photon energy approaches the K-resonance
and can reach $50\%$ already at 3 keV. Therefore an X-ray scattering experiment using intense  XFEL fs-pulses cannot probe the ground state electron density of a crystal. Using XFEL pulses the measured ASF will always be smaller than the form factor measured with conventional synchrotron radiation. The deviation of the measured electron density from the ground state electron density increases for photon energies closer to the K-resonance. However, major changes of diffraction intensity are expected above a certain threshold of pulse fluence. This threshold can be extracted from Fig. \ref{Fig.13.flux} and is supposed to be close to $1.6 $\ mJ/$\mu$m$^2$ using a focus spot of $\sim 1\,\mu$m$^2$. This is remarkable because diffraction is still possible in spite of the fact that this value is much greater than those found in experiment \cite{Hau-Riege, Chalupsky, Hau-Riege-LCLS}. As seen in Fig. \ref{Fig.12.energy} this threshold decreases with decreasing photon energy, and has to be considered using a photon energy close to the K-edge. In an upcoming paper we will investigate the shape of a Bragg peak as a function of the pulse width and photon energy. This is important for the application of solids as optical elements.

\section{Acknowlegements}

This work was supported by the BMBF under grant 05K10PSA. The authors would like to thank Prof. Dr. R. Santra and members of his group for fruitful discussions.

\section*{APPENDIX I: CROSS-SECTIONS CALCULATION}

We have calculated all necessary cross-sections by means of an analytical model of atom with effective charges \cite{FerTriguk} for each shell. The simplest one is the photo-ionization cross-section that can be written in atomic units as \cite{SantraPhoto, LandauV4}

\begin{eqnarray}
\label{22}
    \sigma^{(Ph)}_{nl} (\omega, p) = \frac{2}{3} \frac{\pi \alpha \omega}{p} g_{nl} \sum_{l_j = l \pm 1} \frac{l_>}{2l+1}
    \left| \int_0^{+\infty} r^3 R_{nl}(r) R_{pl_j}(r) dr \right|^2,
\end{eqnarray}

\noindent where $\alpha$ is the fine-structure constant, $\omega$ is the photon frequency, $p=\sqrt{2m(\hbar\omega - \varepsilon_{nl})}$ is the momentum of the photoelectron, and $E_{nl}$ and $g_{nl}$ are the ionization potential and the occupation number of the $(nl)$-subshell, respectively.

Within the framework of ECM, the hydrogen-like wave functions for discrete ($R_{nl}(r)$) and continuous ($R_{pl}(r)$) spectra were used:

\begin{eqnarray}
\label{22a}
    R_{nl}(r) = N_{nl} \left( \frac{2 Z_{nl} r}{n} \right)^l e^{-\frac{Z_{nl} r}{n}} F(-n+l+1,2l+2,\frac{2 Z_{nl} r}{n}),\notag\\
      N_{nl} =\frac{1}{(2l+1)!} \sqrt{\frac{(n+l)!}{2n(n-l-1)!}} \left(\frac{2 Z_{nl}}{n} \right)^\frac{3}{2},\\
    R_{pl}(r) = N_{pl} (2 p r)^l e^{-i p r} F(i \nu + l+1,2l+2,2ipr),\notag\\
      N_{pl} = \frac{Z}{(2l+1)!} \sqrt{\frac{8 \pi}{\nu (1-e^{-2 \pi \nu})}} \prod_{s=1}^l \sqrt{s^2+\nu^2}, \quad \nu \equiv \frac{Z}{p},
\label{22b}
\end{eqnarray}

\noindent with $Z$ and $Z_{nl}$ being the total charge and effective charge of the $(nl)$-subshell, respectively. These values are calculated by a universal formula derived in \cite{FerTriguk}. One should note that for numerical simulation the integrals in (\ref{22}) with the functions (\ref{22a})--(\ref{22b}), and hence the photoionization cross-section, can be calculated analytically.

To calculate the electron-impact ionization cross-section and consider the three-body recombination process we use the binary-encounter dipole model \cite{Kim94, Kai2010} with all ionization potentials being calculated within the framework of  ECM.


\begin{thebibliography}{[1]}



\bibitem{XFEL1}
J. Feldhaus, J. Arthur, and J.B. Hastings, J. Phys. B: At. Mol.
Opt. Phys. \textbf{38}, S799 (2005).

\bibitem{XFEL2}
C. Pellegrini and S. Reiche, J. Sel. Top. Quantum Electron. \textbf{10},
1393 (2004).

\bibitem{Chapman_2009Nat}
H.N. Chapman, Nature Materials \textbf{8}, 299 (2009).

\bibitem{EuroXFEL_tech}
M. Altarelli et al., EuroXFEL Technical Design Report (DESY XFEL Project Group, Hamburg, 2007), http://xfel.desy.de/technical\_information/ .

\bibitem{image}
R. Neutze et al., Nature \textbf{406}, 752 (2000).

\bibitem{Shastri}
S.D. Shastri, P. Zambianchi, and D.M. Mills, J. Synchrotron Rad. \textbf{8}, 1131 (2001).

\bibitem{PSI2009}
Ultrafast Phenomena at the Nanoscale: Science opportunities at the SwissFEL X-ray Laser, Paul Sherrer Institute, September 2009, p. 88.

\bibitem{Hau-Riege}
S.P. Hau-Riege et al., Phys. Rev. Lett. \textbf{98}, 145502 (2007).

\bibitem{Hau-Riege-LCLS}
S.P. Hau-Riege et al., Optics Express \textbf{18},  23936 (2010).

\bibitem{Roseker}
W. Roseker et al., Optics Letters \textbf{34}, 1768--1770 (2009).

\bibitem{Bushuev}
V. Bushuev et al., Proc. of SPIE \textbf{8141}, 81410T-1 (2011).

\bibitem{Shvydko}
Y. Shvyd'ko and R. Lindberg, Phys. Rev. Special Topics---Accelerators and Beams, \textbf{15}, 100702 (2012).

\bibitem{Malgrange}
C. Malgrange and W. Graeff, J. Synchrotron Radiat. \textbf{10}(Pt 3):248--54 (2003).

\bibitem{Hau-Riege-book}
S.P. Hau-Riege, High-Intensity X-rays---Interaction with Matter, Wiley-VCH, Weinheim, Germany, 2011.

\bibitem{Authier}
A. Authier, Dynamical Theory of X-Ray Diffraction, International Union of Crystallography Monographs on Crystallography, 2003.

\bibitem{Santra-Son}
S.-K. Son, L. Young, and R. Santra, Phys. Rev. A. \textbf{83}, 033402 (2011).

\bibitem{SantraPhoto}
R. Santra, J. Phys. B: At. Mol. Opt. Phys. \textbf{42}, 023001 (2009).

\bibitem{ZiajaNew}
B. Ziaja et al., Phys. Rev. B. \textbf{66}, 024116 (2002).

\bibitem{Hau-Riege2013}
S.P. Hau-Riege, Phys. Rev. E. \textbf{87}, 053102 (2013).

\bibitem{Rost2012}
C. Gnodtke, U. Saalmann, and J.-M. Rost, Phys. Rev. Lett. \textbf{108}, 175003 (2012).

\bibitem{Iwayama}
H. Iwayama et al., J. Phys. B \textbf{42}, 134019 (2009).

\bibitem{Bostedt}
C. Bostedt et al., New J. Phys. \textbf{12}, 083004 (2010).

\bibitem{Schorb}
S. Schorb et al., PRL \textbf{108}, 233401 (2012).

\bibitem{Acta}
I.D.~Feranchuk, L.I.~Gurskii, L.I.~Komarov, O.M.~Lugovskaya, F.~Burg\"azy, and A.P.~Ulyanenkov, Acta Cryst. A  \textbf{58}, 370 (2002).

\bibitem{FerTriguk}
V.V. Triguk and I.D. Feranchuk, J. Appl. Spectroscopy \textbf{77}, 749 (2011).

\bibitem{Santra-Ziaja2012}
R. Santra et al., New J. Phys. \textbf{14}, 115015 (2012).

\bibitem{Stepanov}
S.A. Stepanov, X-ray dynamical diffraction web server:
http://sergey.gmca.aps.anl.gov


\bibitem{LandauV4}
L.D. Landau and E.M. Lifshitz, Quantum Electrodynamics. Nauka, Moscow, 1989.

\bibitem{Popov}
V.S. Popov, Uspekhi Fiz. Nauk (in Russian) \textbf{174}, 921 (2004).

\bibitem{solid}
J.M. Ziman, Principles of The Theory of Solids. Cambridge University Press, 1972.

\bibitem{NIST}
NIST Electron Inelastic-Mean-Free-Path Database, http://www.nist.gov/srd/nist71.cfm

\bibitem{Landau8}
L.D. Landau and E.M. Lifshitz, Electrodynamics of Condensed Matter. Nauka, Moscow, 1982.

\bibitem{LANL}
LANL Atomic Physics Codes http://aphysics2.lanl.gov

\bibitem{Batterman}
W. Batterman and H. Cole, Rev. Mod. Phys. \textbf{36}, 681 (1964).

\bibitem{Kissel}
L. Kissel et al., Acta Cryst. \textbf{A51}, 271 (1995).

\bibitem{LandauV10}
L.D. Landau and E.M. Lifshitz, Physical Kinetics. Fizmatlit, Moscow, 2001.

\bibitem{MacRos}
W.M. MacDonald, M.N. Rosenbluth, and W. Chuck, Phys. Rev. \textbf{107}, 350 (1957).

\bibitem{ELENDIF}
W.L. Morgan and B.M. Penetrante, Comp. Phys. Comm. \textbf{58}, 127 (1990).

\bibitem{Vartanyants}
U. Lorenz, N.M. Kabachnik, E. Weckert,  and I.A. Vartanyants, Phys. Rev. E. \textbf{86}, 051911 (2012).

\bibitem{Chalupsky}
J. Chalupsky et al., Appl. Phys. Lett. \textbf{95}, 031111 (2009).

\bibitem{Kim94}
Y.-K. Kim and M.E. Rudd, Phys. Rev. A. \textbf{50}, 3954 (1994).

\bibitem{Kai2010}
T. Kai, Phys. Rev. A. \textbf{81}, 023201 (2010).



\end{thebibliography}
\end{document}